\documentclass[preprint]{aastex631}

\usepackage{csquotes}
\usepackage{graphicx}

\accepted{March 11, 2024}

\shorttitle{Seasonal Variability of Upper Stratospheric Winds on Titan}
\shortauthors{Light et al.}

\graphicspath{{./}{figures/}}

\begin{document}

\title{Measurements of Titan’s Stratospheric Winds during the 2009 Equinox with the eSMA}

\correspondingauthor{Siobhan Light}
\email{slight@umd.edu}

\author[0000-0003-0644-3198]{Siobhan Light}
\affiliation{University of Maryland College Park, 4296 Stadium Drive, College Park, MD 20742, USA}
\affiliation{Planetary Systems Laboratory, NASA Goddard Space Flight Center, Greenbelt, MD 20771, USA}
\affiliation{CRESST, University of Maryland, College Park, MD 20742, USA}

\author[0000-0003-0685-3621]{Mark Gurwell}
\affiliation{Center for Astrophysics $\vert$ Harvard \& Smithsonian, 60 Garden Street, Cambridge, MA 02138, USA}

\author[0000-0002-8178-1042]{Alexander Thelen}
\affiliation{Planetary Systems Laboratory, NASA Goddard Space Flight Center, Greenbelt, MD 20771, USA}
\affiliation{Division of Geological and Planetary Sciences, California Institute of Technology, 1200 East California Boulevard MC 150-21, Pasadena, CA 91125, USA}

\author[0000-0001-8621-6520]{Nicholas A Lombardo}
\affiliation{Department of Earth and Planetary Sciences, Yale University, New Haven, CT, 06511, USA}

\author[0000-0001-9540-9121]{Conor Nixon}
\affiliation{Planetary System Laboratory, NASA Goddard Space Flight Center, Greenbelt, MD 20771, USA}

\begin{abstract}

Saturn's moon Titan possesses stratospheric zonal winds that places it among a sparse class of planetary bodies known to have superrotation in their atmospheres. Few measurements have been made of these speeds in the upper stratosphere, leaving their seasonal variations still not well understood. We examined observations made with the extended Submillimeter Array (eSMA) in 2009 March (L\textsubscript{s}=355\textdegree{}) and 2010 February (L\textsubscript{s}=5\textdegree{}), shortly before and after Titan’s northern spring equinox. Cassini observations and atmospheric models find equinoctial periods to be especially dynamic. Zonal wind calculations, derived from the Doppler frequency shift of CH\textsubscript{3}CN near 349.4 GHz, yielded speeds of 128 $\pm$ 27 m s\textsuperscript{-1} in 2009 and 209 $\pm$ 48 m s\textsuperscript{-1} in 2010. We estimated the measured emission to originate from vertical altitudes of $336^{+112}_{-88}$ kilometers, equivalent to pressures of $3.8^{+19.2}_{-3.4}$ Pa, commensurate with Titan's upper stratosphere/lower mesosphere. This suggests a possible increase in zonal speeds during this period. The results are then compared to those from previous Cassini-inferred and direct-interferometric observations of winds, as well as general circulation model simulations, to form a more complete picture of the seasonal cycle of stratospheric zonal winds.

\end{abstract}


\section{Introduction} \label{sec:intro}

Titan’s atmospheric chemistry surpasses in complexity any other known atmosphere in our solar system \citep{2017JGRE..122..432H}. The atmosphere contains a wealth of organic molecules including hydrocarbons and nitriles \citep[e.g.,][]{2004JGRE..109.6002W}. Over the years, these molecules have been observed to vary both spatially and temporally. Various studies have demonstrated this observationally (see, e.g., \citet{teanby2019seasonal,COUSTENIS2020113413,MATHE2020113547,vinatier2020temperature}) by comparing limb composition measurements taken by Cassini/Composite Infrared Spectrometer (CIRS) of various molecules throughout Titan's latitudes over the course of the mission from 2004 to 2017.

One agent responsible for the fluctuations in Titan’s atmosphere over time is the seasonally changing solar insolation. The length of a Titan year is the same as Saturn’s at 29.5 Earth yr, and Saturn has a similar axial tilt to Earth of about 26.7\textdegree . The changing direction of solar forcing drives a seasonally variable atmosphere on Saturn, and likewise Titan which has a similar tilt to the ecliptic. Seasonal changes were observed with the Cassini mission, but since Cassini was not able to witness a full Titan year or sample all atmospheric levels, more observations are needed to construct a complete picture of the seasonal states of the atmosphere \citep{teanby2019seasonal}.

When it comes to Titan’s general atmospheric circulation, the moon generally has a single pole-to-pole Hadley-like cell, but near the equinoxes, the circulation reverses, resulting in the temporary creation of two equator-to-pole cells \citep{1995Icar..117..358H,2011Icar..213..636N,2012Natur.491..732T,lombardo2023influence}. Titan’s last equinox (2009-2010) occurred on 2009 August 11, and increasing trace gas abundances were observed over Titan’s south pole following this event \citep{2012Natur.491..732T}. Titan’s equinoctial periods appear to be a high activity point for Titan’s meteorology. For example, \citet{2018NatGe..11..727R} reported Cassini observational evidence for active dust storms on Titan during the 2009 and 2010 time period. 

Titan's stratosphere and mesosphere (collectively known as the middle atmosphere) exhibit year-round, strong prograde (west-to-east) zonal winds. These winds are in balance with meridional temperature gradients via the gradient wind relation \citep{sharkey2021potential} and are driven by adiabatic cooling above the polar regions by thermal emission.

One molecule detected numerous times in Titan’s atmosphere due to its prominent spectral character in submillimeter astronomical observations has been acetonitrile (CH\textsubscript{3}CN). This organic nitrile was first observed with the Institut de Radioastronomie Millim{\'e}trique (IRAM) 30 m telescope \citep{1993DPS....25.2509B}. Following this, \citet{2002Icar..158..532M} used IRAM observations of several CH\textsubscript{3}CN transitions to retrieve a disk-averaged vertical profile of the molecule up to 500 km and also demonstrated that CH\textsubscript{3}CN’s distinct lines made it a good candidate molecule to probe the middle atmosphere of Titan. 

In recent years, the Atacama Large Millimeter/submillimeter Array (ALMA) has ushered in a new era of more accurate, direct-wind and molecular-abundance measurements on Titan. These new measurements have increased understanding of the latitudinal variability of acetonitrile \citep[e.g.,][]{2019AJ....158...76C,2019AJ....157..219T}, but its temporal variability over long periods of time remains to be constrained. Furthermore, several years ago ($\sim$L\textsubscript{S}=82\textdegree{}), an unexpectedly intense wind jet incongruent with previous theoretical models was detected with ALMA in the thermosphere (up to 1000 km), though winds were detected all the way from the stratosphere through the mesosphere and up to the thermosphere using different molecules and transitions \citep{2019NatAs...3..614L}. Subsequent work investigated the mechanisms behind the jet, and it highlights how Titan’s atmospheric dynamics are highly temporally variable and altitude dependent \citep{2020ApJ...904L..12C}.

With that in mind, work remains to be done to discern the long-term seasonal variability of Titan’s winds. Doppler measurements of Titan’s winds have been more frequently measured in the ALMA era, but Titan’s long seasonal cycle means that observations made before ALMA remain valuable for addressing this question. In this study, spatially resolved interferometric observations made by the extended Submillimeter Array (eSMA) taken around the time of Titan’s last equinox were analyzed in order to see (1) how CH\textsubscript{3}CN emissions changed in the short period preceding and following Titan’s equinox and (2) the winds on Titan at this time. Both of these atmospheric variables will be examined in the context of previous work to better understand how they have changed on Titan over time.

\section{Observations} \label{sec:Observations}

\begin{figure}
\gridline{\fig{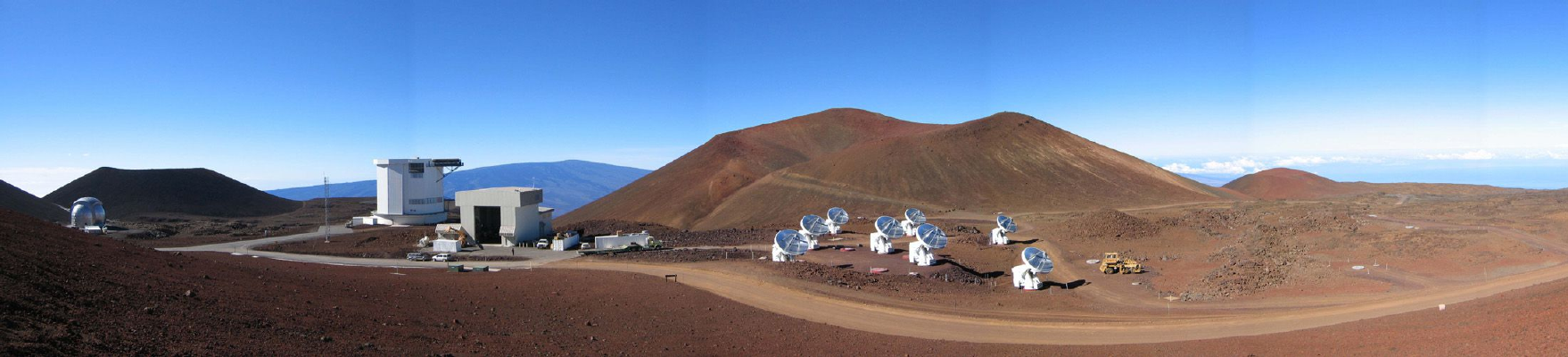}{0.4\textwidth}{(a)}
         }
\gridline{\fig{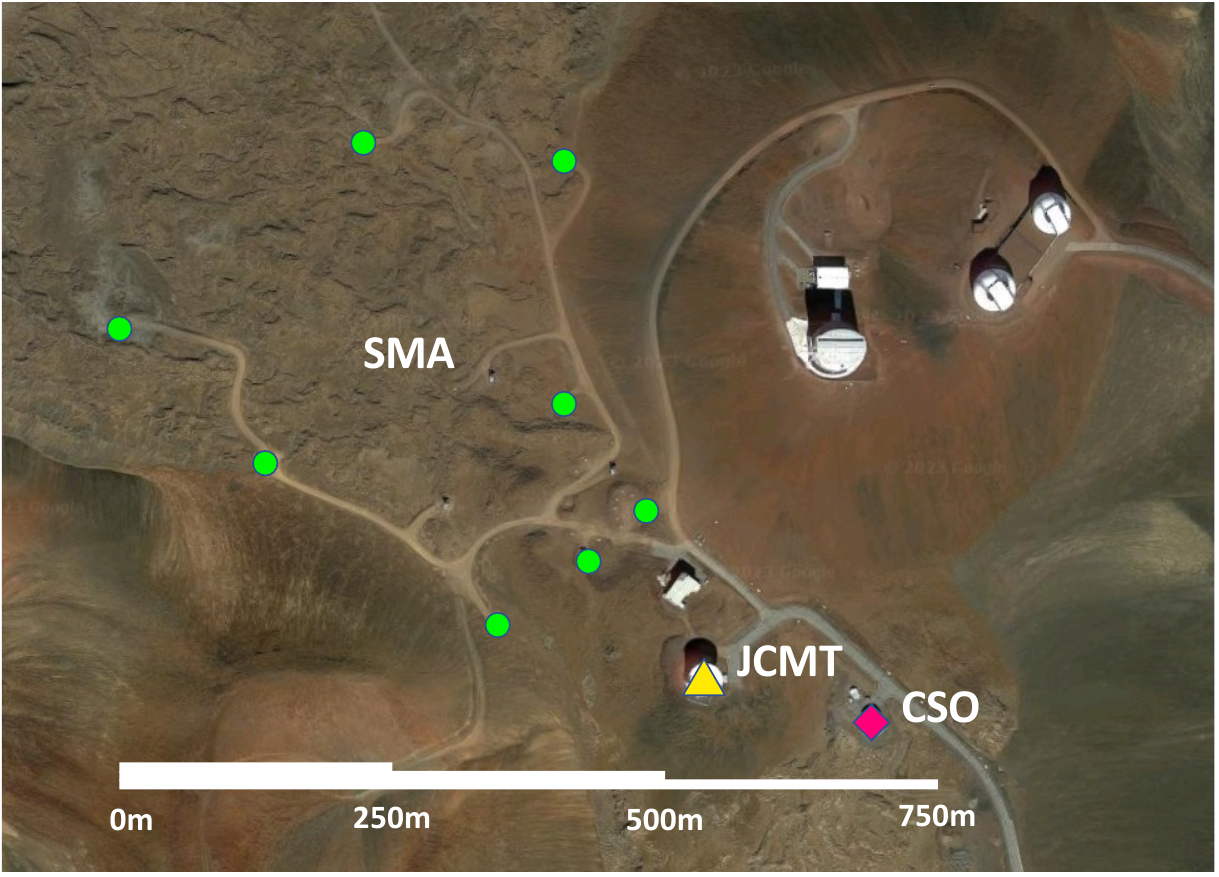}{0.4\textwidth}{(b)}
         }
\caption{(a) Photograph of the observatories involved with the eSMA, left to right: SMA, JCMT, CSO. (Photo Credit: Jonathan Weintroub) (b) Map view; small green circles indicate the location of SMA antennas, yellow triangle represents JCMT location, and the pink diamond represents CSO location. (Photo Credit: Google Earth)}\label{fig:photo} 
\end{figure}

Dedicated interferometric observations of Titan with the eSMA occurred both before and after the Titan equinox, on 2009 March 23  (L\textsubscript{s}=355\textdegree{}) and 2010 February 12  (L\textsubscript{s}=5\textdegree{}), respectively \citep{gurwell2009high,2021LPI....52.2165L}. The Submillimeter Array (SMA), located near the summit of Maunakea, is an array of eight 6 m diameter antennas operating at frequencies between 200 and 420 GHz. The eSMA was a joint project to combine the SMA with two other nearby submillimeter facilities into a larger interferometer, the 15 m diameter James Clerk Maxwell Telescope (JCMT) and the 10.4 m CalTech Submillimeter Observatory (CSO; \citet{2008SPIE.7012E..0DB}; see Figure \ref{fig:photo}).

During the 2009 observations, the eSMA was composed of all eight SMA antennas, the JCMT and the CSO. However, in 2010, only 7 SMA antennas were available, along with the JCMT, providing less overall collecting area and a shorter maximum baseline (see Table \ref{details}). For both sets of observations, the weather was very good. Additional observational details are outlined in Table \ref{details}.

\begin{deluxetable*}{cccccccccc}
\tablecaption{Observational Details\label{details}}
\tablecolumns{10}
\tablenum{1}
\tablewidth{0pt}
\tablehead{
\colhead{Observation Time\tablenotemark{a}} & \colhead{D\tablenotemark{b}} & \colhead{Lat.\tablenotemark{c}} & \colhead{PWV\tablenotemark{d}} & \colhead{$\theta$\tablenotemark{e}} & \colhead{Apparent Size\tablenotemark{f}} & \colhead{Beam Size\tablenotemark{g}} & \colhead{PA\tablenotemark{h}} & \colhead{B\tablenotemark{i}} & \colhead{A\tablenotemark{j}} \\
\colhead{} & \colhead{(au)} & \colhead{(deg)} & \colhead{(mm)} & \colhead{(arcsec)} & \colhead{(arcsec)} & \colhead{(arcsec)} & \colhead{(deg)} & \colhead{(m)} & \colhead{(m\textsuperscript{2})} 
}
\startdata
2009-03-23 06:40-14:25 & 8.4314 & -2.9822 & 0.8 & 182.2 & 0.8422 & 0.384 $\times$ 0.196 & 48.8 & 780 & 488 \\
2010-02-12 09:06-18:30 & 8.7061 & 4.5841 & 1.0 & 120.8 & 0.8156 & 0.382 $\times$ 0.228 & 53.8 & 625 & 375 \\
\enddata
\tablenotemark{a}{UT time} 
\tablenotemark{b}{Distance to Titan} 
\tablenotemark{c}{Subobserver latitude: \url{https://ssd.jpl.nasa.gov/horizons.cgi}} 
\tablenotemark{d}{Precipitable water vapor}
\tablenotemark{e}{Angular  separation between Saturn and Titan, as seen by eSMA}
\tablenotemark{f}{Solid body diameter (5150 km)}
\tablenotemark{g}{Natural beam shape}
\tablenotemark{h}{Position angle (PA) of natural beam}
\tablenotemark{i}{Maximum baseline}
\tablenotemark{j}{Physical collecting area}
\end{deluxetable*}
\vspace{-8.5mm}
In both years, the eSMA was tuned to capture emission from acetonitrile, specifically the CH\textsubscript{3}CN \textit{J} = 19-18 rotational band, near 349.4 GHz. The strongest transitions (k = 0, 1, 2, 3) were measured with $\sim$ 0.203 MHz resolution within a $\sim$91 MHz wide band (448 channels), along with another 1600 MHz wide window measured at 3.25 MHz resolution which provided a continuum signal around the CH\textsubscript{3}CN transitions.

In 2009, there was no real-time Doppler tracking of the changing velocity difference between the observatory and Titan, and alignment of spectra (obtained at 30 s intervals) was performed with offline processing. However, in 2010, real-time Doppler tracking was performed by adjusting the local oscillator (LO) frequency, and offline corrections were not needed.

\subsection{Data Calibration} \label{subsec:calibration}

The interferometric data were calibrated using the Millimeter Interferometer Reduction (MIR) package, a complete suite of tools for calibration of eSMA data, for use in IDL\footnote{\url{https://lweb.cfa.harvard.edu/~cqi/mircook.html}}. The calibration steps were mostly straightforward and standard; we highlight a few important aspects here.

Initial calibration of complex (amplitude and phase) gain corrections with time were measured using periodic observations of a nearby point source, typically millimeter-wave strong blazars; in 2009 the source was J1058+0133 (4C +01.28), and in 2010 J1229+0203 (3C 273). The complex spectral passband response was measured in both years using 3C 273. The signal-to-noise ratio (S/N) of the data (further detailed in the results section) was found to be high enough that self-calibration of complex gains over time could be completed using the established Titan continuum. In doing so, short-term atmospheric instability was removed from the results, further enhancing their quality. Moreover, the flux density scale was set using the Titan continuum, which is accurate to 3-5\% in the submillimeter \citep{butler2012alma}.

\begin{figure}
\includegraphics[width=1\linewidth]{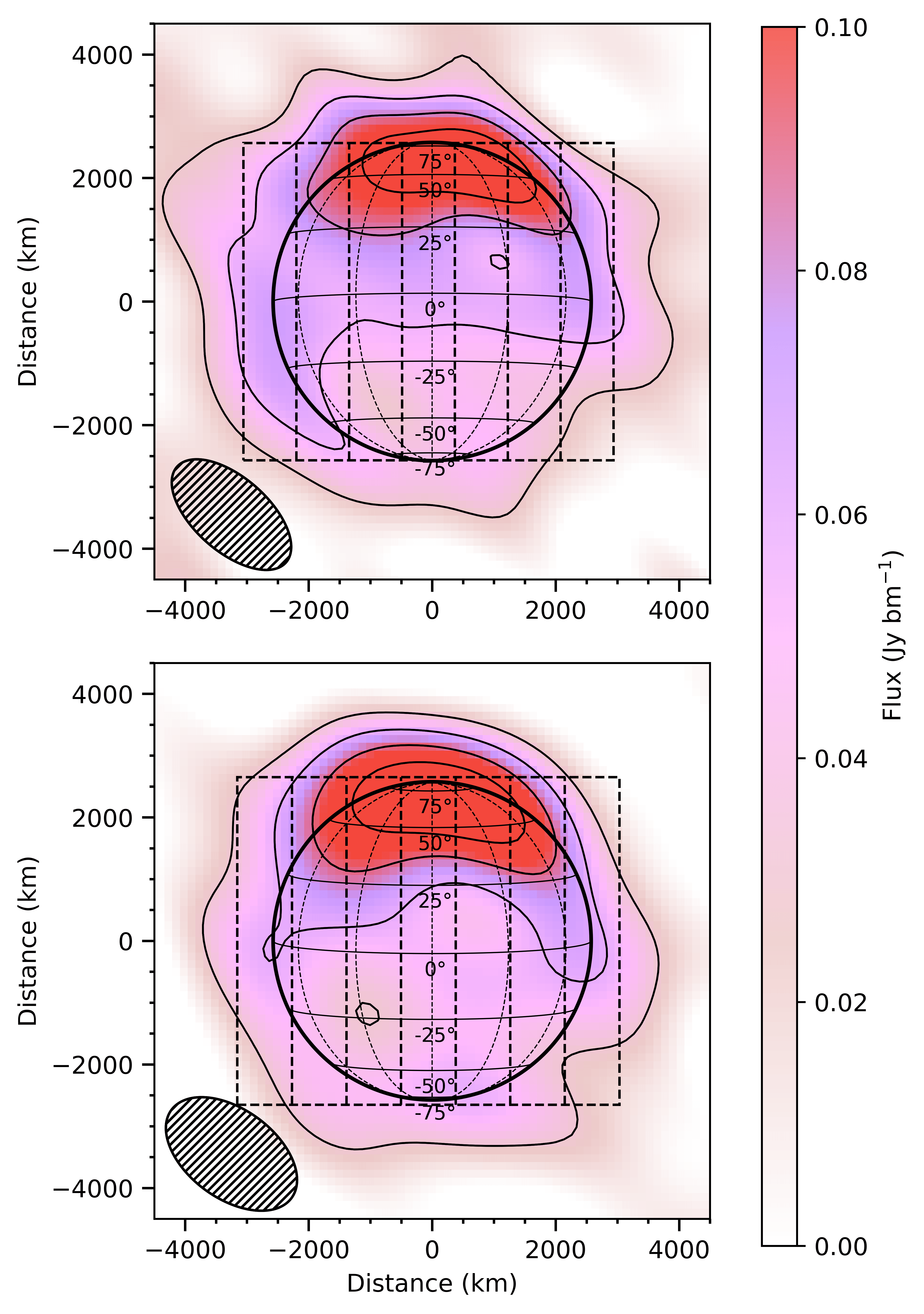}
\caption{Integrated across all channels continuum-subtracted CH\textsubscript{3}CN (\textit{J} = 19 - 18, $\nu$ = 0) centered emission maps for 2009 (top panel) and 2010 (bottom panel). Plotting adapted from \citet{2018ApJ...859L..15C}. Wireframes indicate Titan's surface and orientation in the field of view, with latitudes denoted. The images were convolved to natural beams. The plots show the rectangular areas integrated to calculate zonal wind speed. Each panel shows the beam FWHM (i.e., spatial resolution) in its bottom right corner. Contour intervals are based on percentiles of the maximum flux.} \label{fig:map} 
\end{figure}

The calibrated complex interferometric data (\enquote{visibility data}) of Titan for a broad-frequency continuum band and in each spectral channel were then translated into the UVFITS format. Some further adjustments to the visibility data were applied:
\begin{enumerate}
\item A Doppler correction to each time step of the 2009 visibility data was performed to remove the smearing effects of Earth rotation, Titan's orbital motion, and other velocity variation with time, aligning the spectral correlator coverage at each time step with the Titan frame of reference.
\item The 2010 data set was adjusted such that the apparent diameter of Titan (0.8156") was scaled to the 2009 data (0.8422"), for ease of comparison between the years. This was accomplished by scaling the (u,v) coordinates of each visibility by the ratio of the 2009/2010 diameters, and scaling the amplitude of each visibility by the square of the ratio of the 2009/2010 diameters. This maintains the same 2010 synthesized beam relative to Titan's diameter as well as providing a consistent flux scale and S/N, but directly comparable to the 2009 data.
\item The north pole position angle of Titan for each year was about 356 degrees (measured east of north), amounting to a slight rotation of Titan north relative to J2000 sky north. The north pole was rotated modestly (through rotation of the observed (u,v) positions) such that Titan north and sky north are aligned.
\item For ease of imaging, the Titan continuum data for each visibility point was subtracted from the spectral line data, creating a continuum-subtracted visibility data set; these data represent the spectrum of CH\textsubscript{3}CN relative to the continuum (this manifests entirely as emission in these observations).
\end{enumerate}
The visibility data were then imported into the Astronomical Image Processing System (AIPS). Images of Titan in the continuum and in each spectral channel (continuum-subtracted) were performed using a modified \enquote{clean} (deconvolution) process, using the task IMAGR. The pixel size was set at 0.02". The spectral line and continuum images were both reconvolved to produce images with shapes reflecting the naturally weighted beam (\enquote{natural} beam): a 0.384” $\times$ 0.196” ellipsoid with a position angle (PA) of 48.78$^{\circ}$ for 2009 and 0.382” $\times$ 0.228” with a PA of 53.83$^{\circ}$ for 2010). Via a sigma-clipping function, the mean RMS noise on the integrated flux maps for 2009 and 2010 for the natural beam were measured to be 0.0191 Jy bm$^{-1}$ and 0.0187 Jy bm$^{-1}$, respectively. The final results are shown in the continuum-subtracted maps of Titan shown in Figure \ref{fig:map}.

\section{Results} \label{sec:Results}

\begin{figure*}
\gridline{\fig{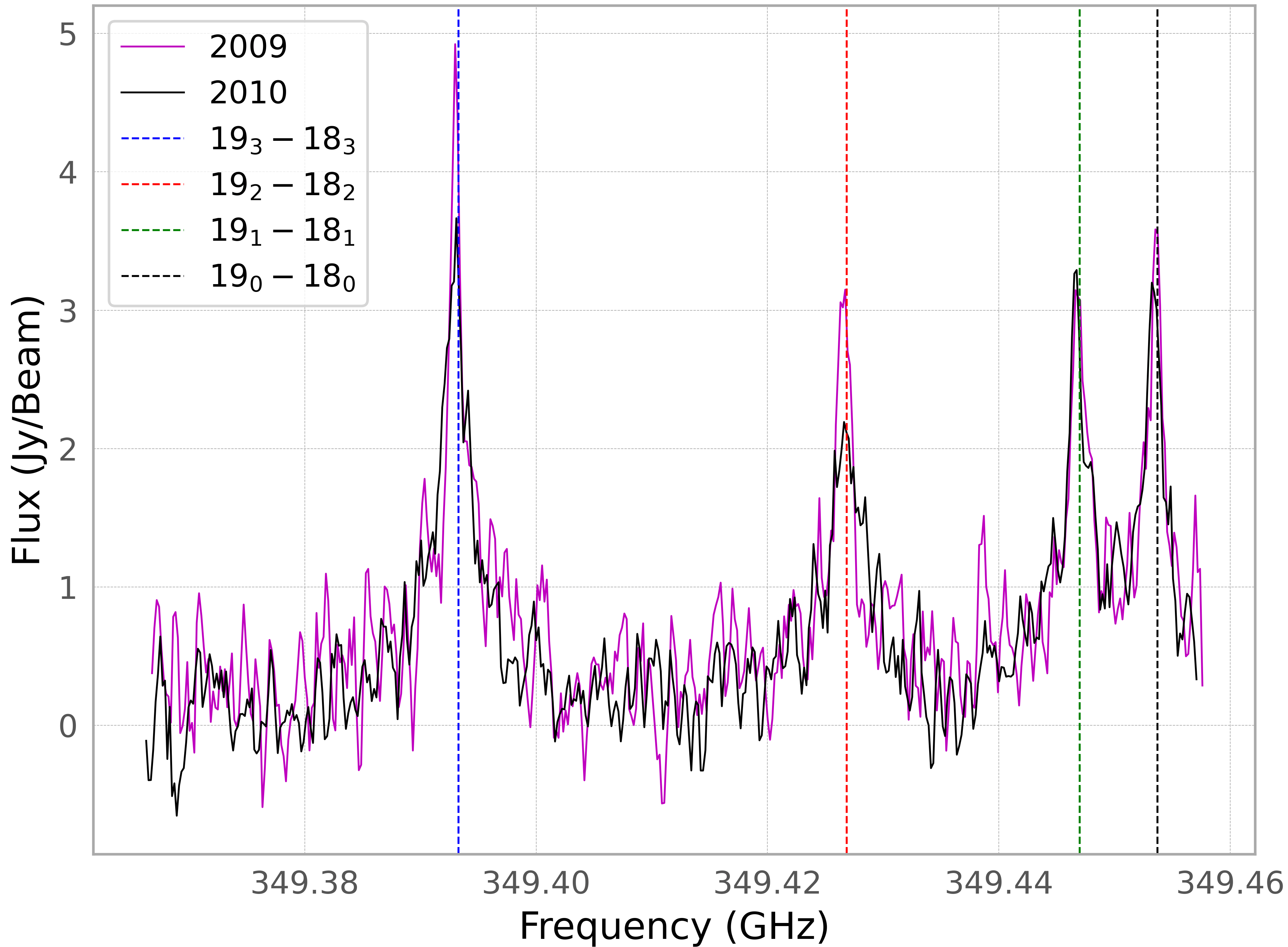}{0.5\textwidth}{(a)}
          \fig{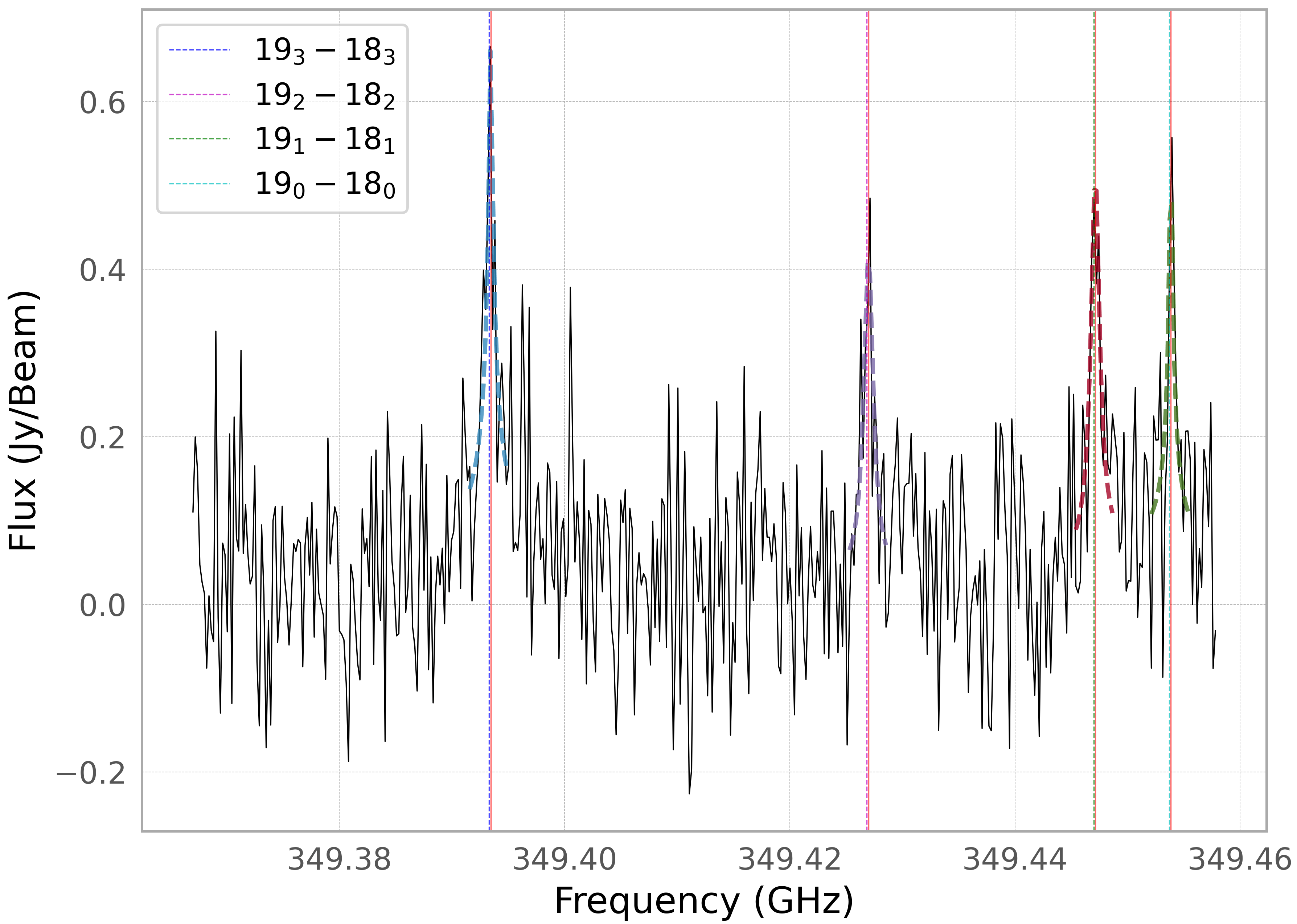}{0.5\textwidth}{(b)}
          }
\gridline{\fig{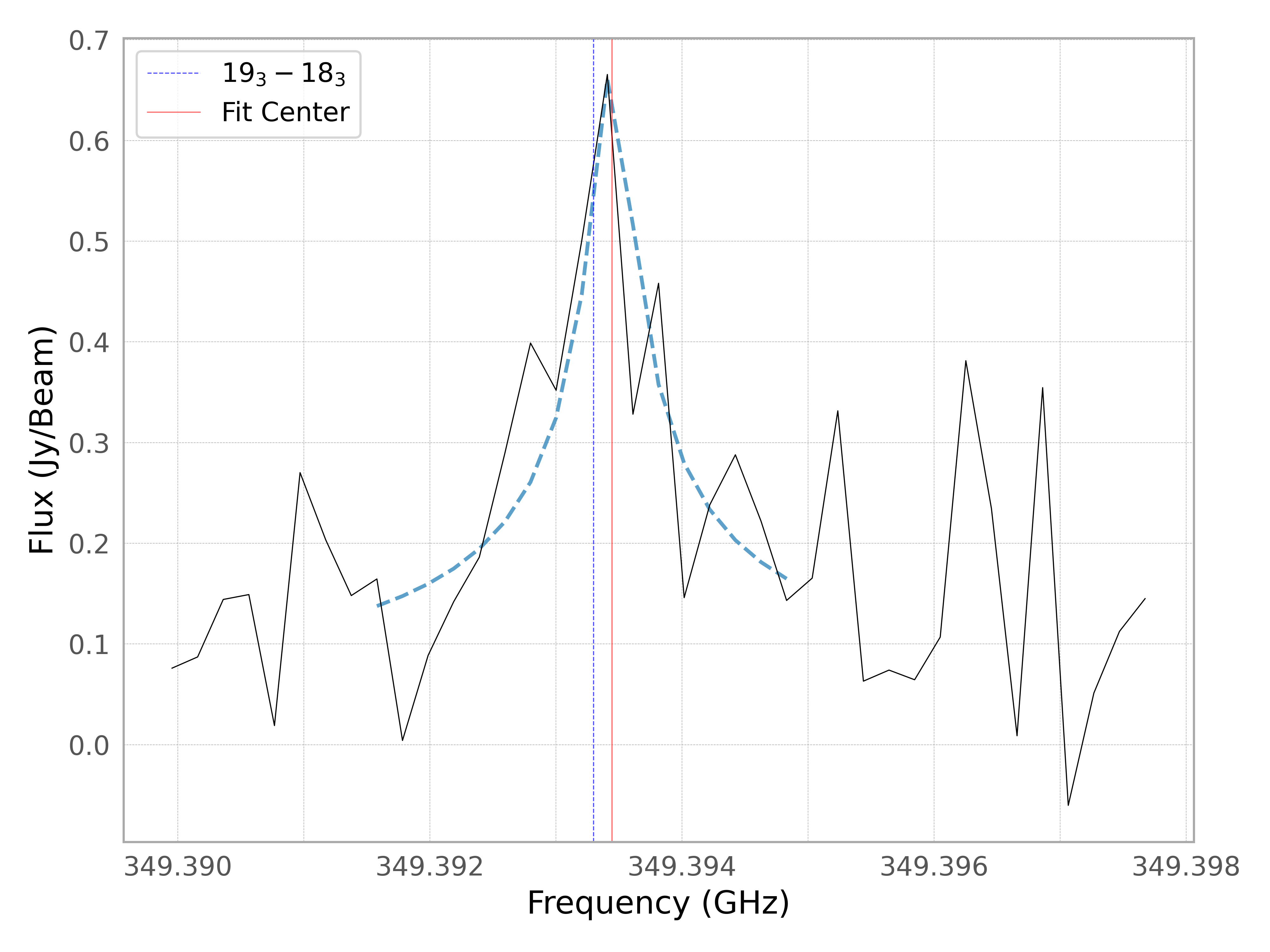}{0.5\textwidth}{(c)}
          \fig{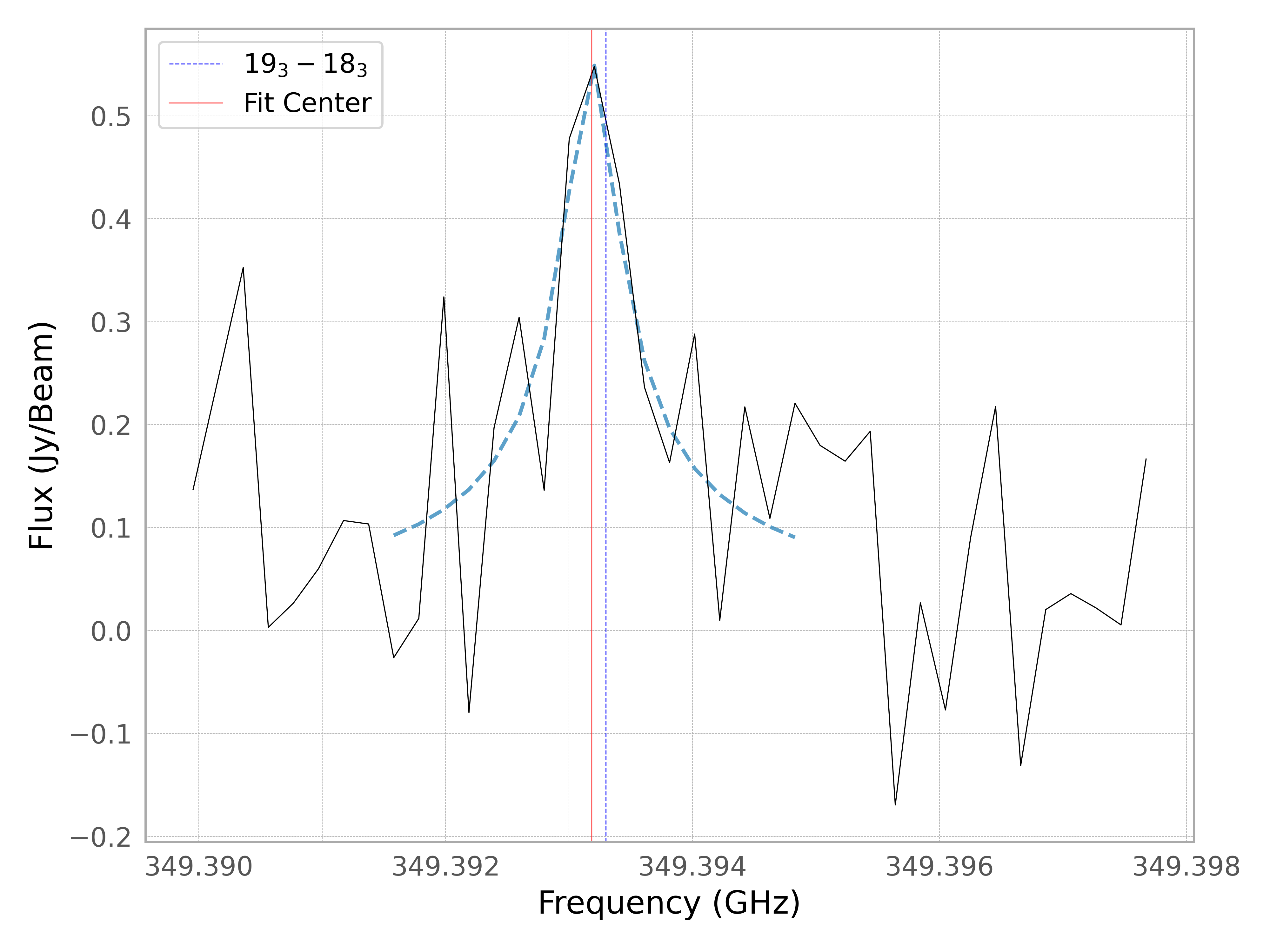}{0.5\textwidth}{(d)}
          }
\caption{(a) Moving average of integrated flux over a radius 0.60" centered on Titan, smoothing together three channels at a time. Each channel has a resolution of 0.203125 MHz. The spectrum from 2009 is in purple and 2010 is in black. Individual transition lines are noted with dashed lines since Doppler shift measurements were made relative to these CH\textsubscript{3}CN \textit{J} = 19\textsubscript{k} - 18\textsubscript{k} (with k=3,2,1,0) transitions. (b) 2009 spectra from the east limb of Titan, with fits overlapped in dashed lines (c) Fit of the 19\textsubscript{3} - 18\textsubscript{3} component of the 2009 eastern limb spectra, with a red line showing the center of the fit; the blue, dotted line shows the rest frequency of the transition, and the turquoise, dashed line shows the model fit (d) Similar to (c) but for the western limb}
\label{fig:spectra}
\end{figure*}

Spectra were extracted from the continuum-subtracted images at frequencies near 349.4 GHz and the results of this over the disk of Titan are shown in Figure \ref{fig:spectra}. The center of the peak emissions was compared to values for the \textit{J} = 19-18 rotational band transitions of CH\textsubscript{3}CN listed in the JPL catalog \citep{pickett1998submillimeter} and Cologne Database for Molecular Spectroscopy (CDMS) \citep{muller2001cologne, muller2005cologne, endres2016cologne}. The two databases effectively agreed on frequency values and associated errors, with the largest error on these values equal to 300 Hz, making it negligible compared to the channel resolution of about 0.203 MHz ($\sim$175 m s\textsuperscript{-1}). 

Typical thermal profiles for Titan's atmosphere \citep{2019NatAs...3..614L} indicate that thermal broadening dominates the line core shape \citep{2020ApJ...904L..12C}. As in \citet{2005AA...437..319M}, \citet{2020ApJ...904L..12C}, and \citet{ 2019NatAs...3..614L}, pressure-broadened line wings also were observed in the line emission. After fitting various line shapes to the spectral peaks, a Moffat distribution \citep{moffat1969theoretical} was found to more accurately represent the complex line emission from Titan than either Gaussian, Lorentzian, or Voigt line shapes, and was consequently used for determining the Doppler shifts of the observed line shapes, as shown in Figure \ref{fig:spectra}. The equation for the Moffat distribution is
\begin{equation}
 f(x; A, \mu, \sigma, \beta) = A[(\frac{x-\mu}{\sigma}^{2})+1]^{-\beta}
\end{equation}
with four parameters: amplitude ($A$), center ($\mu$), width ($\sigma$), and exponent ($\beta$). The Moffat distribution was able to model both the line core and wings of the CH\textsubscript{3}CN more precisely than other profile types.

To measure the Doppler shift and determine zonal winds, the calculations were performed using seven rectangular regions taken from across the disk of Titan, highlighted in Figure \ref{fig:map}. The reason large areas needed to have their emissions integrated together to resolve zonal winds is because the interferometric observations described here, with natural beamwidths approximated as elliptical Gaussians of the FWHM $\sim$0.38" $\times$ 0.20", only coarsely resolve the disk of Titan ($\sim$0.8" diameter, see Table \ref{details}). Half of the integrated flux of a Gaussian beam comes from outside the central FWHM, meaning that emission from well outside the central region of the beam contributes to the image at a given pixel. This is especially relevant for these CH\textsubscript{3}CN observations, where the emission (even at the coarse resolution) is strongly concentrated at the limb of Titan as seen in the integrated beam maps (Figure \ref{fig:map}). This is a result of submillimeter lines being generally optically thin \citep[e.g.][]{2020ApJ...904L..12C} and presenting the strongest thermal emission where path lengths in the atmosphere are largest (i.e., at the limbs). The combination of relatively large beam size and highly concentrated emission means that careful consideration of the beam shape is important. For example, spectra from the disk center show apparent emission, but this is largely limb emission captured by the coarse beam (mostly along the long axis of the beam), not emission from the disk center. The rotational model described below takes this effect into account.

Spectra across latitudes in each of the rectangular regions were then summed together and adjusted by a geometric scale factor determined by
\begin{equation}
S=\frac{2\pi ab}{\Delta l^{2} \log{2}}
\end{equation}
where $a$ and $b$ are the beam major and minor axis respectively and $\Delta$\textit{l} is pixel size. After including continuum error, the spectra were fit at each of the four transitions using the LMFIT package for Python \citep{newville2016lmfit}. The resulting fits for the easternmost region are shown in Figure \ref{fig:spectra}b for the 2009 data. The shift between limbs within the single \textit{J} = 19-18 rotational band is expected to be essentially the same since they probe the same altitude range (see \citet{2005AA...437..319M}). Calculations performed in Subsection \ref{subsec:abundance} also support this choice. Consequently, due to the challenges of interpreting data in this low S/N regime, sensitivity is enhanced by calculating a single frequency shift by finding a weighted average of the individual transition shifts. These frequencies were then converted to velocities via the Doppler shift formula.

\begin{figure}
    \centering
    \resizebox{\columnwidth}{!}{%
        \includegraphics{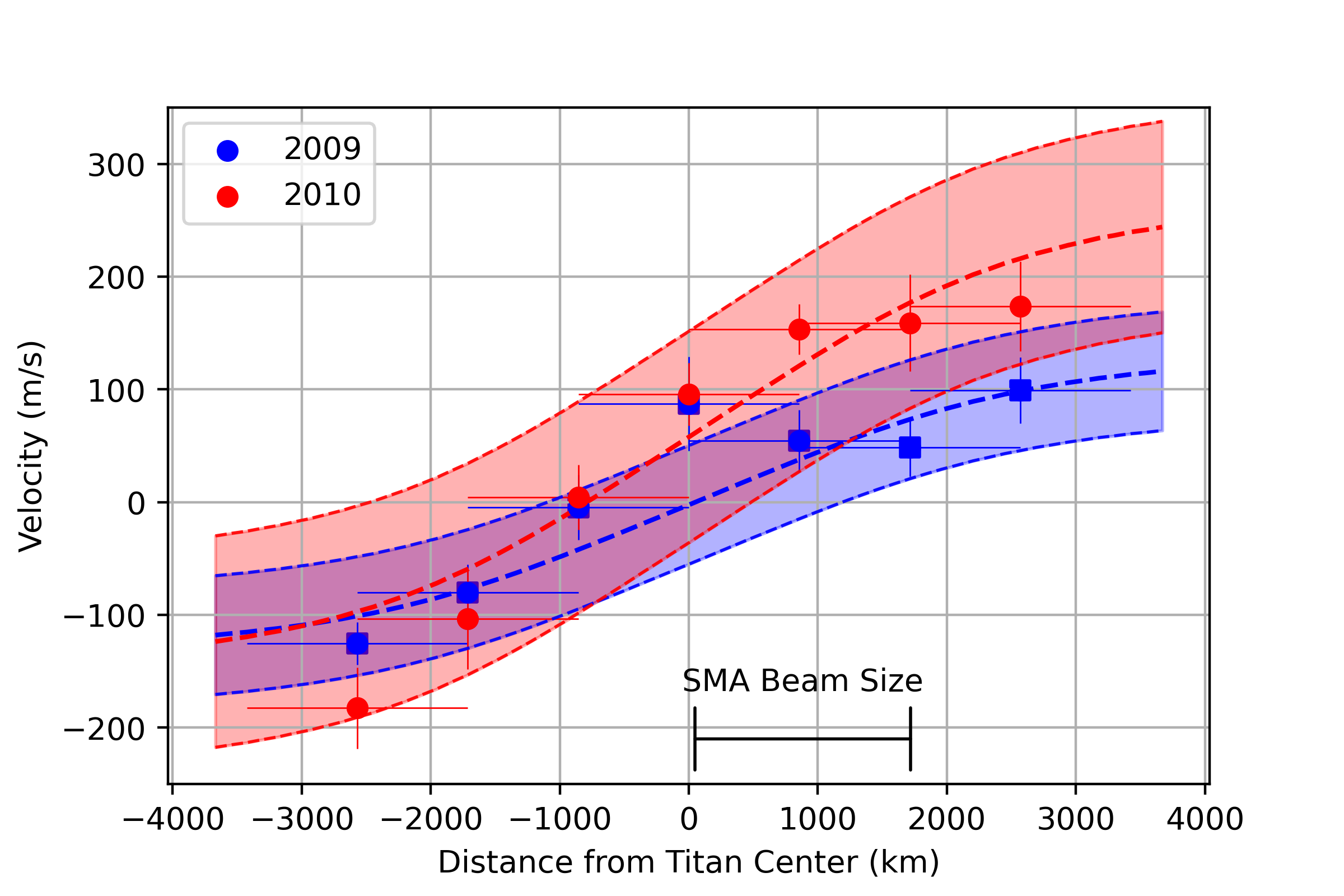}%
    }
    \caption{Comparison of the measured Doppler shifted velocities from seven rectangular regions spanning all latitudes on Titan, along with their associated errors. Blue squares represent 2009 data, while red circles depict 2010 data. Dashed lines, optimized to align with solid body rotation acting as a proxy for zonal wind, are fit for each respective year. Shaded regions indicate the 95\% confidence interval ($\pm$1.96$\sigma$) on the velocity fits. The SMA beam size corresponds to the east-west projection of the major axis of the 2009 \enquote{natural} beam shape.}
    \label{fig:vel}
\end{figure}

To calculate uncertainty on each regional velocity, the same fitting method was applied to 20,000 sets of synthetic Gaussian data based on the spectral resolution, noise, and shape of the original spectral data. The subsequent synthetic data fitting results were filtered to exclude velocities that were outside 4$\sigma$ so that extreme synthetic data results did not affect the final error calculations. To determine how wide a spectral region to fit over, the fitting methodology and subsequent error determination was tested on regions of 1.015 MHz to 3.453 MHz, limited on the lower end by the spectral FWHM and thermal broadening and on the higher end by the proximity of the $19_{1}-18_{1}$ and $19_{0}-18_{0}$ transitions. After doing so, the root-mean-square errors (RMSE) of the predicted shifts in the synthetic data sets was used to estimate the error on the velocity measurements. The lowest errors and most consistent velocity values were associated with using $\pm$ 1.625 MHz (equivalent to $\pm$ 8 channels of resolution). Figure \ref{fig:vel} displays velocity values from each of the seven rectangular regions, represented for 2009 as blue squares and 2010 as red squares, and their associated error bars.

As mentioned above, the large beam size means that the measured velocity will be offset relative to the true equatorial east-west zonal wind. To account for this, a simple model was made based on a fully zonal wind approximating solid body rotation. Zonal wind speed scales as a function of latitude as V\textsubscript{z}cos($\phi$), where V\textsubscript{z} is the equatorial maximum zonal wind speed and $\phi$ is latitude. Approximation of a roughly solid body zonal wind model is appropriate here, considering the relatively large synthesized beam, that the results of \citet{2019NatAs...3..614L} and \citet{2020ApJ...904L..12C} found a strong zonal wind system at lower latitudes, and that for a viewing geometry with a sub-Earth latitude of zero, 81.8$\%$ of the apparent disk is from $\pm$45 deg (\enquote{low}) latitudes. This distribution was used to generate a smeared velocity map, allowing beam effects to be accounted for when calculating zonal velocity. However, this relied on the assumption that smearing the calculated zonal wind radial velocity by weighting the measured emission and the beam yields the same result. The model also captures how the relative velocity of CH\textsubscript{3}CN emission is averaged within each of the seven rectangular regions (centered on Titan) are affected by the distribution of emission. As the emission maps in Figure \ref{fig:map} show, the model also assumes emission is weaker towards the south pole and stronger towards the north pole. The velocity values in each rectangular region are hence adjusted to reflect the relative velocity computed by the model. Subsequently, we utilize a curve fitting approach to model Titan's solid body rotation, shown in Figure \ref{fig:vel}. The fitting algorithm uses \texttt{scipy.optimize} module from the SciPy library \citep{2020SciPy-NMeth} to minimize using the Nelder-Mead algorithm \citep{gao2012implementing} the weighted sum of squared differences differences between the observed velocities and those predicted by the velocity curve. To accommodate potential Doppler tracking errors, we allow for flexibility in the "zero" velocity, rather than fixing it. Thus, the model we were optimize fitting was
\begin{equation}
V = (Z*r) + \omega
\end{equation}
where $V$ is the predicted velocity, $Z$ denotes the global zonal wind value, $r$ represents the relative velocity of the zonal wind at a given distance in the solid body rotation model, and $\omega$ signifies the velocity offset at the center of Titan. Essentially, we are fitting the one-dimensional rectangular velocity data to a solid body rotation curve that acts as a proxy for global zonal winds. Furthermore, we subtract the rotation speed of Titan at the approximate stratospheric altitude of the data, as it is not relevant to the observed zonal winds.

Directly fitting the data yields an estimate of approximately 138 m s\textsuperscript{-1} and 249 m s\textsuperscript{-1} for the global zonal winds in 2009 and 2010, respectively. However, uncertainties exist in both the distances and velocities associated with each data point, compounded by the limited number of data points available for fitting. To address this, we generated two million sets of velocities and distances, assuming a random normal distribution associated with each point's average distance and measured velocity. Subsequently, these synthetic datasets were fitted repeatedly using the same method, allowing us to derive the average and standard deviation of the synthetic data model parameters. These statistics were then used to calculate the global zonal winds and associated uncertainties.

By this method, the zonal wind speed for 2009 was calculated to be 128 $\pm$ 27 m s\textsuperscript{-1} with an offset of -0.4 m s\textsuperscript{-1} $\pm$ 16 m s\textsuperscript{-1}. In 2010, the zonal wind speed increased to 209 $\pm$ 48 m s\textsuperscript{-1} with a larger offset of $\sim$61 m s\textsuperscript{-1} $\pm$ 27 m s\textsuperscript{-1}. We found that the wind strengthened by a factor of 1.6 $\pm$ 0.5, with a 3.2 $\sigma$ confidence. Overall, the weighted average zonal wind speed for this 2009-2010 time period is 169 $\pm$ 27 m s\textsuperscript{-1}.

\subsection{Altitude Calculation}
\label{subsec:abundance}

\begin{figure*}
    \centering
    \resizebox{\linewidth}{!}{%
        \includegraphics{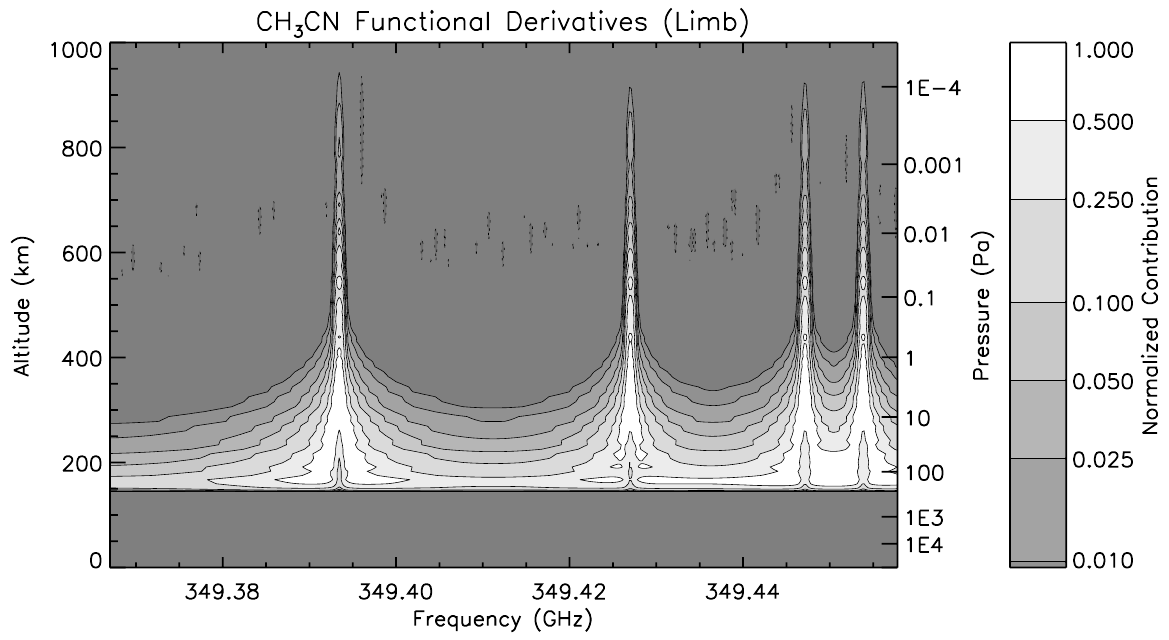}%
    }
    \caption{Normalized functional derivatives calculated using NEMESIS of the spectral radiance with respect to 2009 CH\textsubscript{3}CN (\textit{J} = 19 - 18, $\nu$ = 0) abundance, pressure/altitude, and frequency. Absolute contour values are shown in side bar (e.g., the white contour shade corresponds to the highest contribution, $\sim$1.000).}
    \label{fig:profiles}
\end{figure*}

Using the Non-linear optimal Estimator for MultivariatE Spectral analySIS (NEMESIS) radiative transfer code \citep{irwin2008nemesis} and similar methods to \citet{thelen2019abundance}, we calculated relevant functional derivatives from 150 to 1200 kilometers in altitude for a spectrum extracted with the \enquote{natural} 2009 beam shape at the Eastern limb. CH\textsubscript{3}CN spectral line parameters were incorporated from the Cologne Database for Molecular Spectroscopy (CDMS) \citep{muller2001cologne, muller2005cologne, endres2016cologne} and the HITRAN catalog \citep{gordon2022hitran2020}. N\textsubscript{2}-broadening parameters were adopted from \citet{dudaryonok2015n2}. In turn, this enabled us to estimate the altitude and pressure levels to which the subsequent wind measurements were primarily derived from. Figure \ref{fig:profiles} shows the normalized contribution from an altitude of $\sim$150 to $\sim$1000 kilometers since contributions below or above these altitudes are negligible. The functional derivatives (and thus the extracted spectra) are not sensitive to altitudes below 150 km for reasons outlined in \citet{thelen2019abundance} based on calculations made from 3 \textit{a priori} profiles \citep{2002Icar..158..532M, dobrijevic2018photochemical, loison2015neutral} demonstrating that CH\textsubscript{3}CN emission at these frequencies is not sensitive to gas abundances at these low altitudes.

Taking the average of the peak contributions at the four emission line peaks, and using the same weights used to later calculate the Doppler shift, the emission appears to have originated approximately at vertical altitudes of $336^{+112}_{-88}$ kilometers corresponding to pressures of $3.8^{+19.2}_{-3.4}$ Pa. These altitudes correspond to Titan’s upper stratosphere/lower mesosphere and are similar to previous studies involving CH\textsubscript{3}CN \citep[e.g.][]{2005AA...437..319M, 2019NatAs...3..614L, thelen2019abundance, 2019AJ....158...76C}). Although there are secondary peaks in the upper atmosphere, the contribution is orders of magnitudes less than the stratosphere/mesosphere region. Despite seasonal changes in latitudinal concentration, since the molecule has a long photochemical lifetime in the middle atmosphere \citep{wilson2004current, loison2015neutral,2019Icar..324..120V} vertical abundances likely do not strongly vary over the course of a Earth year. In particular, according to \citet{2019Icar..324..120V}, CH\textsubscript{3}CN has a photochemical lifetime of about 91 years at 300 kilometers. From 2012 to 2015, there were also no major differences between the northern and southern hemispheres with regard to upper stratosphere CH\textsubscript{3}CN abundances \citep{thelen2019abundance}. Further work in \citet{2019AJ....158...76C} concluded the molecule appears to be relatively well-mixed at these stratospheric altitudes compared to shorter-lived nitriles and hydrocarbons.

\section{Discussion} \label{sec:Discussion}

Cassini observations around this time period \citep{teanby2010seasonal} observed trace gas abundance to be enriched at the north pole and depleted at the south pole. The emission maps in Figure \ref{fig:map} are consistent with these observations and show that little significant change occurred in CH\textsubscript{3}CN emissions between 2009 and 2010. Acetonitrile remained concentrated primarily in northern latitudes during these observations, similar to the 2013 ALMA observations described in \citet{thelen2019abundance}, supporting the idea that confinement by the northern winter polar vortex persisted after the equinox \citep{teanby2019seasonal} and due to the relatively long photochemical lifetime of CH\textsubscript{3}CN in Titan's stratosphere. Close to the second set of observations reported here, Cassini CIRS in January 2010 observed polar warming at high altitudes soon after equinox but no large increases in abundances until approximately a year later \citep{2012Natur.491..732T}. There appears to be a slightly higher concentration of acetonitrile in the east of Titan prior to the equinox than after, although noise in the data makes this an uncertain conclusion. Initial analysis of the 2009 data found an increasing CH\textsubscript{3}CN abundance with altitude \citep{gurwell2009high}. 

\begin{deluxetable*}{ccccc}
\tablecaption{Stratospheric Wind Speed Measurements\label{compare}} 
\tablecolumns{5}
\tablenum{2}
\tablewidth{0pt}
\tablehead{
\colhead{Work} & \colhead{Observational Year} & \colhead{L\textsubscript{s} (\textdegree{})} & \colhead{$\bar{z (km)}$} & \colhead{v\textsubscript{zonal} (m s\textsuperscript{-1})}}
\startdata
{\cite{hubbard1993occultation}} & 1989 & 88 & 350$\pm$100 & $\sim$100\textsuperscript{*} \\
{\cite{2001GeoRL..28.2361K}} & 1993-1996 & 163-195 & 110-300 & 210$\pm$150 \\
{\cite{kostiuk2005titan}} & 2003 & 300 & 130-300 & 190$\pm$90 \\
{\cite{sicardy2006two}} & 2003 & 311 & $\sim$250 & $\sim$100-150 \\  
{\cite{2005AA...437..319M}} & 2003-2004 & 257-315 & 300$\pm$150 & 160$\pm$60 \\
This Work & 2009 & 355 & $336^{+112}_{-88}$ & 128$\pm$27 \\
This Work & 2010 & 5 & $336^{+112}_{-88}$ & 209$\pm$48 \\
{\cite{2019NatAs...3..614L}} & 2016 & 82 & $345^{+60}_{-140}$ & 220 \\
{\cite{2020ApJ...904L..12C}} & 2016 & 82 & $\sim$345 & 254$\pm$4 \\
{\cite{2020ApJ...904L..12C}} & 2017 & 90 & $\sim$345 & 185$\pm$3 \\
\enddata
\tablecomments{\textsuperscript{*}\footnotesize{Order of magnitude}}
\end{deluxetable*}
\vspace{-8.5mm}
Table \ref{compare} shows how the zonal wind measurements in this study compare to others made with similar Doppler shift methodology at comparable stratospheric altitudes \citep{hubbard1993occultation, 2001GeoRL..28.2361K, sicardy2006two, kostiuk2005titan, 2005AA...437..319M, 2019NatAs...3..614L, 2020ApJ...904L..12C}. Consistent with all of these previous observations, the 2009 and 2010 eSMA measurements show a prograde zonal wind flow and support the conclusion of superrotation of Titan’s middle atmosphere. 

To contextualize Titan's seasons, its northern summer solstice (L\textsubscript{s}=90\textdegree{}) occurred in February 1988 and May 2017, northern fall equinox (L\textsubscript{s}=180\textdegree{}) in November 1995 (next in 2025), northern winter solstice (L\textsubscript{s}=270\textdegree{}) in November 2002, and northern spring equinox (L\textsubscript{s}=0\textdegree{}) in August 2009. The first direct wind measurements were derived from data taken during a 1989 occultation and noise precluded measurements being known more precisely than to an order of magnitude of around $\sim$100 m s\textsuperscript{-1} \citep{hubbard1993occultation}. \citet{2001GeoRL..28.2361K} combined observations from August 1993, October 1995, and September 1996 to derive speeds of 210$\pm$150 m s\textsuperscript{-1} peaking at around 210 kilometers in altitude (notably lower than the level probed in this study). \cite{kostiuk2005titan} and \cite{sicardy2006two} also probed lower in the stratosphere, deriving winds of 190$\pm$90 and $\sim$100-150 m s\textsuperscript{-1}, respectively.

Our measurements from 2009 and 2010 suggest a strengthening of Titan's stratospheric zonal winds over its equinox.  The strength of the 2009 zonal winds are comparable to those observed half a season prior during northern winter in 2003/2004 \citep{2005AA...437..319M}, and are significantly weaker than those measured shortly before summer solstice in 2016 \citep{2019NatAs...3..614L, 2020ApJ...904L..12C}. The measurements before the northern spring equinox from the eSMA do appear to be lower than those taken at the summer solstice in 2017 \citep{2020ApJ...904L..12C} while the measurements taken after the equinox in 2010 are more similar in value. The results from 2009 generally suggest that the equinoctial period may be one of the slowest points in the upper stratospheric jet found to date. \citet{2020ApJ...904L..12C} suggested that the fastest zonal jets may occur in the years leading up to Titan’s solstices and slowing until equinoxes. The results presented here remain consistent with this idea and further suggest that the time period before Titan’s northern spring equinox may be a low point in the speed of stratospheric zonal jets. Moreover, the overall 2010 zonal wind speed result appears to be closer in magnitude to those derived shortly before the 2016 summer solstice as reported in \citet{2019NatAs...3..614L} and \citet{2020ApJ...904L..12C}. This suggests a rapid increase in the speed of the upper stratospheric jet shortly after the northern equinox. However, more measurements must be made across Titan’s seasonal cycle and at different altitudes (particularly the mesosphere and thermosphere) to definitively conclude the seasonal variability of Titan’s zonal jets. For example, it would take additional data points in the time period between these 2010 observations and the subsequent 2016 observations to say whether the stratospheric jet maintained its speed through Titan’s spring, or whether its speed varied over the course of northern spring.

General circulation models (GCMs) are a useful tool in simulating and diagnosing large scale climate systems in planetary atmospheres. Over the years, several general circulation models (GCMs) of Titan’s atmosphere have been developed \citep{1995Icar..117..358H,2011Icar..213..636N,lebonnois2012titan,lora2015gcm}. Generally, GCMs suggest that zonal winds weaken during winter until hemispheric reversal \citep{2011Icar..213..636N,lora2015gcm}, an idea the low-speed measurement in 2009 would support. Around equinox, upwelling at lower latitudes, closer to the region examined in this study, is suggested by multiple models to bring eastward angular momentum to the upper stratosphere \citep{2011Icar..213..636N,lombardo2023influence}. During equinox, \citet{lombardo2023influence} observed that the dominant zonal jet shifts from the previously winter hemisphere to the hemisphere entering fall. The 2010 results would support their suggestion that zonal winds reach a high value following the equinox. However, we do see a general increase in disk averaged zonal winds from just before to just after equinox. We compared our results to those of TitanWRF \citep{2011Icar..213..636N} and the Titan Atmospheric Model (TAM) \citep{lora2015gcm} to attempt to constrain the timeline of the changes in the middle atmosphere zonal winds near equinox.

\begin{figure*}
    \centering
    \resizebox{\linewidth}{!}{%
        \includegraphics{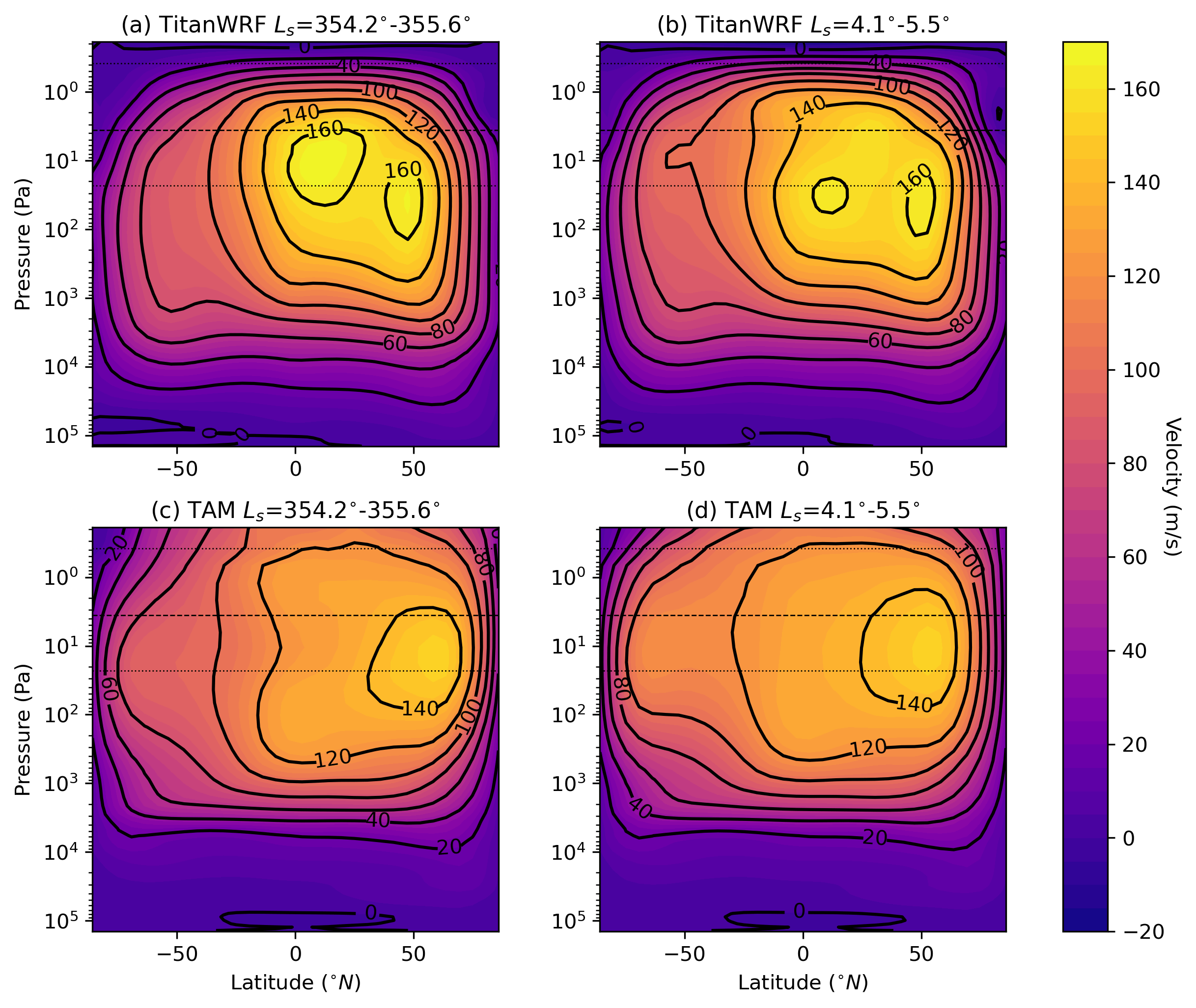}%
    }
    \caption{(a) Simulations from \citet{2011Icar..213..636N} TitanWRF for the 2009 zonal wind velocities (b) Simulations from \citet{2011Icar..213..636N} TitanWRF for the 2010 zonal wind velocities (c) Simulations from \citet{lombardo2023influence} TAM for the 2009 zonal wind velocities (d) Simulations from \citet{lombardo2023influence} TAM for the 2010 zonal wind velocities. Dashed horizontal lines represent the average peak contribution pressures for the eSMA wind measurements, and dotted horizontal lines represent the lower and upper bounds based on the associated error bars, indicating the range of uncertainty around the average peak contribution pressures.}
    \label{fig:gcm}
\end{figure*}

Recently, \citet{shultis2022winter} used  TitanWRF simulations from \citet{2011Icar..213..636N} to constrain the dynamics of Titan’s winter stratosphere, a time period the results presented here complement. This model was found to be in general agreement with CIRS observations at pressures above 1 Pa, enabling direct comparisons to be made. Overall, the model tended to have higher simulated temperatures relative to observations but captured general observed temperature and zonal wind trends. In general, the model simulated a minimum in mid-latitude stratospheric zonal winds between winter solstice and spring equinox, a result potentially complementing the low speed observed with the eSMA in 2009. In Figure \ref{fig:gcm}a, the simulated zonal winds of the TitanWRF GCM at the time of the 2009 observations are shown. In this time period shortly before the equinox, this GCM simulates at comparable altitudes a peak zonal speed of $\sim$150~m s\textsuperscript{-1} in the northern hemisphere 
($\sim$17.5\textdegree{}) and equatorial zonal wind speed of about $\sim$145~m s\textsuperscript{-1}. Little appears to directly change by the time of the 2010 observations, with the GCM simulations of this time period shown in Figure \ref{fig:gcm}b. With regards to the seasonal trends of zonal winds, \citet{shultis2022winter} simulates for both hemispheres around 10 Pa the winds weaken and move towards low latitudes during fall and reach a minimum around mid-winter. To help understand why this happens, this complements how CIRS observed that the winter pole for the upper stratosphere/lower mesosphere was warmer than lower latitudes likely caused by adiabatic heating of the descending air of meridional circulation \citep{achterberg2008titan,achterberg2011temporal,teanby2019seasonal,vinatier2020temperature,sharkey2021potential,shultis2022winter,achterberg2023temporal}. Following this mid-winter weakening, TitanWRF suggests that the zonal winds strengthen slightly and shift poleward just before each hemisphere's respective spring equinox. Following spring equinox, zonal winds are generally thought to increase, corresponding to upper stratosphere angular momentum reaching a maximum \citep{2011Icar..213..636N}. This is due to cooling from less adiabatic compression/warming from less descending air, which in turn led to a steeper meridional temperature gradient, driving stronger winds. Overall, these results support the increase in zonal winds after equinox, but suggest it may occur more rapidly than the TitanWRF simulation appears to predict. Relative to the southern hemisphere, winter in the northern hemisphere tends to drive stronger general circulation, with asymmetries between the two driven by differences in insolation caused by Titan’s orbital eccentricity. As a result, we would expect increases in comparable low-latitude zonal wind speeds following the upcoming 2025 equinox, albeit not as large.

TAM was recently updated to include seasonally varying radiative species  \citep{lombardo2023influence}, which tend to cool the polar latitudes due to increased longwave emission from molecular enrichment. The polar cooling also leads to slightly increased stratospheric zonal wind speeds.  Like TitanWRF, TAM simulates a degree of seasonal symmetry between hemispheres. Figure \ref{fig:gcm}c and \ref{fig:gcm}d show TAM zonal wind simulations for the 2009 and 2010 observation times, respectively. Relative to TitanWRF, TAM simulated less change across this time period than TitanWRF, but TAM suggests zonal winds are in the process of strengthening at the time of the observations, peaking part-way through the northern spring (after the 2010 observations). Within the equatorial region examined in this study (roughly latitudes of 30\textdegree{}S--30\textdegree{}N), TAM simulations show that zonal winds appear to peak shortly after equinoxes and reach minimums shortly after solstices. This seasonal variation is more extreme at the upper stratosphere/lower mesosphere level ($\sim$3 Pa) than lower in the stratosphere ($\sim$50 Pa) \citep{lombardo2023influence}. The eSMA observations also suggest the winds are strengthening, but at a more rapid pace than predicted by TAM. In connection with temperature, this variable appears to generally peak around northern winter solstice followed by an uneven decrease until northern summer solstice. Around northern spring equinox, there appears to be a small temperature dip \citep{https://doi.org/10.1029/2023JE008061}. 

One way to see how the GCMs simulate changes in zonal wind over this study's time period is to calculate from their model outputs a disk averaged (considering area via cosine weighting) zonal profile. Calculating the cosine weighted disk average of the zonal winds modeled by TitanWRF as shown in Figures \ref{fig:gcm}a and \ref{fig:gcm}b, corresponding to the timing of the 2009 and 2010 observations, respectively, reveals that TitanWRF simulated a 10\% decrease in the overall zonal winds between these two time periods, unlike what the observations suggest. Meanwhile, taking the cosine weighted disk averages of Figures \ref{fig:gcm}c and \ref{fig:gcm}d finds TAM simulated a roughly 8\% increase in the zonal winds between the two observation periods. The general increase in zonal winds suggested by the eSMA results during the observed time period would therefore be consistent with TAM simulations, but appears to be of a larger magnitude.

During the same time period as the measurements made here, Cassini observations saw zonal winds in the northern hemisphere gradually decrease as northern winter progressed towards summer, while in the southern hemisphere zonal winds began to increase rapidly after equinox \citep{sharkey2021potential}. Overall, higher zonal wind values around L\textsubscript{s} = 0\textdegree{} would have been driven by higher values in the northern hemisphere. The limited spatial resolution of the eSMA data makes differentiating between the northern and southern hemisphere difficult. The higher zonal wind speeds around equinox includes changes in the wind speeds from both the northern and southern hemispheres. This means the strengthening we observed shortly after equinox likely includes strengthening in the southern hemisphere and weakening in the northern hemisphere, similar to the southern mid-latitude strengthening and slight weakening observed in \citep{sharkey2021potential}. However, the thermal wind measurements they reported left winds unconstrained equatorward of about 20 degrees in either hemisphere at the altitudes we are sensitive to, so we are unable to make comparisons as to the finer low latitude meridonal structure of the zonal winds around equinox.

\section{Conclusion} \label{sec:Conclusion}

New zonal wind measurements of Titan’s upper stratosphere shortly before and after the moon’s northern spring equinox were made using eSMA observations and Doppler shift techniques. The results appear consistent with the idea that zonal wind speeds decrease during winter on Titan followed by a rapid increase in speed following the equinox and may help inform future GCMs to better understand Titan’s atmosphere. Although the spatial resolution of these eSMA measurements are much more coarse than those offered by ALMA in recent years, they nonetheless offer insight into this dynamical time period in Titan’s atmosphere. Continued measurements of Titan’s atmosphere with ALMA and other sub-millimeter observatories are still needed to form a more complete picture of the seasonal variation in the moon’s circulation. For example, knowledge of interannual variability is currently limited to models, so it can not be concluded if the seasonal patterns described here remain constant over future Titan years. By understanding the dynamics of Titan’s atmospheric superrotation and circulation, insights have previously been made into the atmosphere of Venus and may also help to understand those of exoplanets as Doppler observations begin to be made of hot Jupiters \citep{imamura2020superrotation}. Improvements in knowledge of Titan’s middle atmosphere will complement atmospheric modeling that may support the upcoming \textit{Dragonfly} mission design \citep{lorenz2021selection, lorenz2021engineering}.

\section{Acknowledgments} \label{sec:Acknowledgements}

The material is based upon work supported by NASA under award number 80GSFC21M0002. SLL and CAN received funding for this work from the NASA Solar System Observations (SSO) Program and the NASA Astrobiology Institute. The development of the eSMA was facilitated by grant 614.061.416 from the Netherlands Organization for Scientific Research, NWO, and NSF grant AST0540882 to the Caltech Submillimeter Observatory, along with the dedicated work of observatory staff from the JCMT, CSO, and the SMA. The Submillimeter Array is a joint project between the Smithsonian Astrophysical Observatory and the Academia Sinica Institute of Astronomy and Astrophysics and is funded by the Smithsonian Institution and the Academia Sinica. We recognize that Maunakea is a culturally important site for the indigenous Hawaiian people; we are privileged to study the cosmos from its summit. Thank you to Erin Flowers insight into TitanWRF. Thank you to Martin Cordiner for help with the code to plot the acetonitrile emissions and fruitful discussions about previous work related to Titan's zonal winds.

\bibliography{bib}{}

\begin{thebibliography}{}
\expandafter\ifx\csname natexlab\endcsname\relax\def\natexlab#1{#1}\fi
\providecommand{\url}[1]{\href{#1}{#1}}
\providecommand{\dodoi}[1]{doi:~\href{http://doi.org/#1}{\nolinkurl{#1}}}
\providecommand{\doeprint}[1]{\href{http://ascl.net/#1}{\nolinkurl{http://ascl.net/#1}}}
\providecommand{\doarXiv}[1]{\href{https://arxiv.org/abs/#1}{\nolinkurl{https://arxiv.org/abs/#1}}}

\bibitem[{Achterberg(2023)}]{achterberg2023temporal}
Achterberg, R.~K. 2023, The Planetary Science Journal, 4, 140,
  \dodoi{10.3847/PSJ/acebea}

\bibitem[{Achterberg {et~al.}(2008)Achterberg, Conrath, Gierasch, Flasar, \&
  Nixon}]{achterberg2008titan}
Achterberg, R.~K., Conrath, B.~J., Gierasch, P.~J., Flasar, F.~M., \& Nixon,
  C.~A. 2008, Icarus, 194, 263, \dodoi{10.1016/j.icarus.2007.09.029}

\bibitem[{Achterberg {et~al.}(2011)Achterberg, Gierasch, Conrath, Flasar, \&
  Nixon}]{achterberg2011temporal}
Achterberg, R.~K., Gierasch, P.~J., Conrath, B.~J., Flasar, F.~M., \& Nixon,
  C.~A. 2011, Icarus, 211, 686, \dodoi{10.1016/j.icarus.2010.08.009}

\bibitem[{{B{\'e}zard} {et~al.}(1993){B{\'e}zard}, {Marten}, \&
  {Paubert}}]{1993DPS....25.2509B}
{B{\'e}zard}, B., {Marten}, A., \& {Paubert}, G. 1993, in AAS/Division for
  Planetary Sciences Meeting Abstracts, Vol.~25, AAS/Division for Planetary
  Sciences Meeting Abstracts \#25, 25.09

\bibitem[{{Bottinelli} {et~al.}(2008){Bottinelli}, {Young}, {Chamberlin},
  {Tilanus}, {Gurwell}, {Wilner}, {Shinnaga}, {Yoshida}, {Friberg}, {van
  Langevelde}, {van Dishoeck}, {Hogerheijde}, {Hughes}, {Christensen}, {Hills},
  {Richer}, \& {Curtis}}]{2008SPIE.7012E..0DB}
{Bottinelli}, S., {Young}, K.~H., {Chamberlin}, R., {et~al.} 2008, in Society
  of Photo-Optical Instrumentation Engineers (SPIE) Conference Series, Vol.
  7012, Ground-based and Airborne Telescopes II, ed. L.~M. {Stepp} \&
  R.~{Gilmozzi}, 70120D, \dodoi{10.1117/12.788949}

\bibitem[{Butler(2012)}]{butler2012alma}
Butler, B. 2012, ALMA Memos, 594.
\newblock \url{https://library.nrao.edu/public/memos/alma/memo594.pdf}

\bibitem[{{Cordiner} {et~al.}(2020){Cordiner}, {Garcia-Berrios}, {Cosentino},
  {Teanby}, {Newman}, {Nixon}, {Thelen}, \& {Charnley}}]{2020ApJ...904L..12C}
{Cordiner}, M.~A., {Garcia-Berrios}, E., {Cosentino}, R.~G., {et~al.} 2020,
  \apjl, 904, L12, \dodoi{10.3847/2041-8213/abc688}

\bibitem[{{Cordiner} {et~al.}(2018){Cordiner}, {Nixon}, {Charnley}, {Teanby},
  {Molter}, {Kisiel}, \& {Vuitton}}]{2018ApJ...859L..15C}
{Cordiner}, M.~A., {Nixon}, C.~A., {Charnley}, S.~B., {et~al.} 2018, \apjl,
  859, L15, \dodoi{10.3847/2041-8213/aac38d}

\bibitem[{{Cordiner} {et~al.}(2019){Cordiner}, {Teanby}, {Nixon}, {Vuitton},
  {Thelen}, \& {Charnley}}]{2019AJ....158...76C}
{Cordiner}, M.~A., {Teanby}, N.~A., {Nixon}, C.~A., {et~al.} 2019, \aj, 158,
  76, \dodoi{10.3847/1538-3881/ab2d20}

\bibitem[{Coustenis {et~al.}(2020)Coustenis, Jennings, Achterberg, Lavvas,
  Bampasidis, Nixon, \& Flasar}]{COUSTENIS2020113413}
Coustenis, A., Jennings, D., Achterberg, R., {et~al.} 2020, Icarus, 344,
  113413, \dodoi{https://doi.org/10.1016/j.icarus.2019.113413}

\bibitem[{Dobrijevic \& Loison(2018)}]{dobrijevic2018photochemical}
Dobrijevic, M., \& Loison, J. 2018, Icarus, 307, 371,
  \dodoi{10.1016/j.icarus.2017.10.027}

\bibitem[{Dudaryonok {et~al.}(2015)Dudaryonok, Lavrentieva, \&
  Buldyreva}]{dudaryonok2015n2}
Dudaryonok, A.~S., Lavrentieva, N.~N., \& Buldyreva, J.~V. 2015, Icarus, 256,
  30, \dodoi{10.1016/j.icarus.2015.04.025}

\bibitem[{Endres {et~al.}(2016)Endres, Schlemmer, Schilke, Stutzki, \&
  M{\"u}ller}]{endres2016cologne}
Endres, C.~P., Schlemmer, S., Schilke, P., Stutzki, J., \& M{\"u}ller, H.~S.
  2016, Journal of Molecular Spectroscopy, 327, 95,
  \dodoi{10.1016/j.jms.2016.03.005}

\bibitem[{Gao \& Han(2012)}]{gao2012implementing}
Gao, F., \& Han, L. 2012, Computational Optimization and Applications, 51, 259,
  \dodoi{10.1007/s10589-010-9329-3}

\bibitem[{Gordon {et~al.}(2022)Gordon, Rothman, Hargreaves, Hashemi, Karlovets,
  Skinner, Conway, Hill, Kochanov, Tan, {et~al.}}]{gordon2022hitran2020}
Gordon, I.~E., Rothman, L.~S., Hargreaves, R., {et~al.} 2022, Journal of
  quantitative spectroscopy and radiative transfer, 277, 107949,
  \dodoi{10.1016/j.jqsrt.2021.107949}

\bibitem[{Gurwell {et~al.}(2009)Gurwell, Moullet, {et~al.}}]{gurwell2009high}
Gurwell, M.~A., Moullet, A., {et~al.} 2009, in AAS/Division for Planetary
  Sciences Meeting Abstracts\# 41, Vol.~41, 30--07

\bibitem[{{H{\"o}rst}(2017)}]{2017JGRE..122..432H}
{H{\"o}rst}, S.~M. 2017, Journal of Geophysical Research (Planets), 122, 432,
  \dodoi{10.1002/2016JE005240}

\bibitem[{{Hourdin} {et~al.}(1995){Hourdin}, {Talagrand}, {Sadourny},
  {Courtin}, {Gautier}, \& {Mckay}}]{1995Icar..117..358H}
{Hourdin}, F., {Talagrand}, O., {Sadourny}, R., {et~al.} 1995, \icarus, 117,
  358, \dodoi{10.1006/icar.1995.1162}

\bibitem[{Hubbard {et~al.}(1993)Hubbard, Sicardy, Miles, Hollis, Forrest,
  Nicolson, Appleby, Beisker, Bittner, Bode, {et~al.}}]{hubbard1993occultation}
Hubbard, W., Sicardy, B., Miles, R., {et~al.} 1993, Astronomy and Astrophysics,
  269, 541

\bibitem[{Imamura {et~al.}(2020)Imamura, Mitchell, Lebonnois, Kaspi, Showman,
  \& Korablev}]{imamura2020superrotation}
Imamura, T., Mitchell, J., Lebonnois, S., {et~al.} 2020, Space Science Reviews,
  216, 1, \dodoi{10.1007/s11214-020-00703-9}

\bibitem[{Irwin {et~al.}(2008)Irwin, Teanby, De~Kok, Fletcher, Howett, Tsang,
  Wilson, Calcutt, Nixon, \& Parrish}]{irwin2008nemesis}
Irwin, P., Teanby, N., De~Kok, R., {et~al.} 2008, Journal of Quantitative
  Spectroscopy and Radiative Transfer, 109, 1136,
  \dodoi{10.1016/j.jqsrt.2007.11.006}

\bibitem[{{Kostiuk} {et~al.}(2001){Kostiuk}, {Fast}, {Livengood}, {Hewagama},
  {Goldstein}, {Espenak}, \& {Buhl}}]{2001GeoRL..28.2361K}
{Kostiuk}, T., {Fast}, K.~E., {Livengood}, T.~A., {et~al.} 2001, \grl, 28,
  2361, \dodoi{10.1029/2000GL012617}

\bibitem[{Kostiuk {et~al.}(2005)Kostiuk, Livengood, Hewagama, Sonnabend, Fast,
  Murakawa, Tokunaga, Annen, Buhl, \& Schm{\"u}lling}]{kostiuk2005titan}
Kostiuk, T., Livengood, T., Hewagama, T., {et~al.} 2005, Geophysical research
  letters, 32, \dodoi{10.1029/2005GL023897}

\bibitem[{Lebonnois {et~al.}(2012)Lebonnois, Burgalat, Rannou, \&
  Charnay}]{lebonnois2012titan}
Lebonnois, S., Burgalat, J., Rannou, P., \& Charnay, B. 2012, Icarus, 218, 707,
  \dodoi{https://doi.org/10.1016/j.icarus.2011.11.032}

\bibitem[{{Lellouch} {et~al.}(2019){Lellouch}, {Gurwell}, {Moreno}, {Vinatier},
  {Strobel}, {Moullet}, {Butler}, {Lara}, {Hidayat}, \&
  {Villard}}]{2019NatAs...3..614L}
{Lellouch}, E., {Gurwell}, M.~A., {Moreno}, R., {et~al.} 2019, Nature
  Astronomy, 3, 614, \dodoi{10.1038/s41550-019-0749-4}

\bibitem[{{Light} {et~al.}(2021){Light}, {Gurwell}, {Nixon}, \&
  {Thelen}}]{2021LPI....52.2165L}
{Light}, S.~L., {Gurwell}, M.~A., {Nixon}, C.~A., \& {Thelen}, A.~E. 2021, in
  52nd Lunar and Planetary Science Conference, Lunar and Planetary Science
  Conference, 2165

\bibitem[{Loison {et~al.}(2015)Loison, H{\'e}brard, Dobrijevic, Hickson,
  Caralp, Hue, Gronoff, Venot, \& B{\'e}nilan}]{loison2015neutral}
Loison, J., H{\'e}brard, E., Dobrijevic, M., {et~al.} 2015, Icarus, 247, 218,
  \dodoi{10.1016/j.icarus.2014.09.039}

\bibitem[{Lombardo \& Lora(2023{\natexlab{a}})}]{lombardo2023influence}
Lombardo, N.~A., \& Lora, J.~M. 2023{\natexlab{a}}, Icarus, 390, 115291,
  \dodoi{10.1016/j.icarus.2022.115291}

\bibitem[{Lombardo \&
  Lora(2023{\natexlab{b}})}]{https://doi.org/10.1029/2023JE008061}
---. 2023{\natexlab{b}}, Journal of Geophysical Research: Planets, 128,
  e2023JE008061, \dodoi{https://doi.org/10.1029/2023JE008061}

\bibitem[{Lora {et~al.}(2015)Lora, Lunine, \& Russell}]{lora2015gcm}
Lora, J.~M., Lunine, J.~I., \& Russell, J.~L. 2015, Icarus, 250, 516,
  \dodoi{10.1016/j.icarus.2014.12.030}

\bibitem[{Lorenz(2021)}]{lorenz2021engineering}
Lorenz, R.~D. 2021, Advances in Space Research, 67, 2219,
  \dodoi{10.1016/j.asr.2021.01.023}

\bibitem[{Lorenz {et~al.}(2021)Lorenz, MacKenzie, Neish, Le~Gall, Turtle,
  Barnes, Trainer, Werynski, Hedgepeth, \& Karkoschka}]{lorenz2021selection}
Lorenz, R.~D., MacKenzie, S.~M., Neish, C.~D., {et~al.} 2021, The Planetary
  Science Journal, 2, 24, \dodoi{10.3847/PSJ/abd08f}

\bibitem[{{Marten} {et~al.}(2002){Marten}, {Hidayat}, {Biraud}, \&
  {Moreno}}]{2002Icar..158..532M}
{Marten}, A., {Hidayat}, T., {Biraud}, Y., \& {Moreno}, R. 2002, \icarus, 158,
  532, \dodoi{10.1006/icar.2002.6897}

\bibitem[{Mathé {et~al.}(2020)Mathé, Vinatier, Bézard, Lebonnois, Gorius,
  Jennings, Mamoutkine, Guandique, \& {Vatant d’Ollone}}]{MATHE2020113547}
Mathé, C., Vinatier, S., Bézard, B., {et~al.} 2020, Icarus, 344, 113547,
  \dodoi{10.1016/j.icarus.2019.113547}

\bibitem[{Moffat(1969)}]{moffat1969theoretical}
Moffat, A. 1969, Astronomy and Astrophysics, 3, 455

\bibitem[{{Moreno} {et~al.}(2005){Moreno}, {Marten}, \&
  {Hidayat}}]{2005AA...437..319M}
{Moreno}, R., {Marten}, A., \& {Hidayat}, T. 2005, \aap, 437, 319,
  \dodoi{10.1051/0004-6361:20042117}

\bibitem[{M{\"u}ller {et~al.}(2005)M{\"u}ller, Schl{\"o}der, Stutzki, \&
  Winnewisser}]{muller2005cologne}
M{\"u}ller, H.~S., Schl{\"o}der, F., Stutzki, J., \& Winnewisser, G. 2005,
  Journal of Molecular Structure, 742, 215,
  \dodoi{10.1016/j.molstruc.2005.01.027}

\bibitem[{M{\"u}ller {et~al.}(2001)M{\"u}ller, Thorwirth, Roth, \&
  Winnewisser}]{muller2001cologne}
M{\"u}ller, H.~S., Thorwirth, S., Roth, D., \& Winnewisser, G. 2001, Astronomy
  \& Astrophysics, 370, L49, \dodoi{10.1051/0004-6361:20010367}

\bibitem[{{Newman} {et~al.}(2011){Newman}, {Lee}, {Lian}, {Richardson}, \&
  {Toigo}}]{2011Icar..213..636N}
{Newman}, C.~E., {Lee}, C., {Lian}, Y., {Richardson}, M.~I., \& {Toigo}, A.~D.
  2011, \icarus, 213, 636, \dodoi{10.1016/j.icarus.2011.03.025}

\bibitem[{Newville {et~al.}(2016)Newville, Stensitzki, Allen, Rawlik,
  Ingargiola, \& Nelson}]{newville2016lmfit}
Newville, M., Stensitzki, T., Allen, D.~B., {et~al.} 2016, Astrophysics Source
  Code Library, ascl, \dodoi{10.5281/zenodo.598352}

\bibitem[{Pickett {et~al.}(1998)Pickett, Poynter, Cohen, Delitsky, Pearson, \&
  M{\"u}ller}]{pickett1998submillimeter}
Pickett, H., Poynter, R., Cohen, E., {et~al.} 1998, Journal of Quantitative
  Spectroscopy and Radiative Transfer, 60, 883,
  \dodoi{10.1016/S0022-4073(98)00091-0}

\bibitem[{{Rodriguez} {et~al.}(2018){Rodriguez}, {Le Mou{\'e}lic}, {Barnes},
  {Kok}, {Rafkin}, {Lorenz}, {Charnay}, {Radebaugh}, {Narteau}, {Cornet},
  {Bourgeois}, {Lucas}, {Rannou}, {Griffith}, {Coustenis}, {App{\'e}r{\'e}},
  {Hirtzig}, {Sotin}, {Soderblom}, {Brown}, {Bow}, {Vixie}, {Maltagliati},
  {Courrech du Pont}, {Jaumann}, {Stephan}, {Baines}, {Buratti}, {Clark}, \&
  {Nicholson}}]{2018NatGe..11..727R}
{Rodriguez}, S., {Le Mou{\'e}lic}, S., {Barnes}, J.~W., {et~al.} 2018, Nature
  Geoscience, 11, 727, \dodoi{10.1038/s41561-018-0233-2}

\bibitem[{Sharkey {et~al.}(2021)Sharkey, Teanby, Sylvestre, Mitchell, Seviour,
  Nixon, \& Irwin}]{sharkey2021potential}
Sharkey, J., Teanby, N.~A., Sylvestre, M., {et~al.} 2021, Icarus, 354, 114030,
  \dodoi{10.1016/j.icarus.2020.114030}

\bibitem[{Shultis {et~al.}(2022)Shultis, Waugh, Toigo, Newman, Teanby, \&
  Sharkey}]{shultis2022winter}
Shultis, J., Waugh, D., Toigo, A., {et~al.} 2022, The Planetary Science
  Journal, 3, 73, \dodoi{10.3847/PSJ/ac5ea1}

\bibitem[{Sicardy {et~al.}(2006)Sicardy, Colas, Widemann, Bellucci, Beisker,
  Kretlow, Ferri, Lacour, Lecacheux, Lellouch, {et~al.}}]{sicardy2006two}
Sicardy, B., Colas, F., Widemann, T., {et~al.} 2006, Journal of Geophysical
  Research: Planets, 111, \dodoi{10.1029/2005JE002624}

\bibitem[{Teanby {et~al.}(2010)Teanby, Irwin, De~Kok, \&
  Nixon}]{teanby2010seasonal}
Teanby, N., Irwin, P., De~Kok, R., \& Nixon, C. 2010, The Astrophysical Journal
  Letters, 724, L84, \dodoi{10.1088/2041-8205/724/1/L84}

\bibitem[{Teanby {et~al.}(2019)Teanby, Sylvestre, Sharkey, Nixon, Vinatier, \&
  Irwin}]{teanby2019seasonal}
Teanby, N., Sylvestre, M., Sharkey, J., {et~al.} 2019, Geophysical research
  letters, 46, 3079, \dodoi{10.1029/2018GL081401}

\bibitem[{{Teanby} {et~al.}(2012){Teanby}, {Irwin}, {Nixon}, {de Kok},
  {Vinatier}, {Coustenis}, {Sefton-Nash}, {Calcutt}, \&
  {Flasar}}]{2012Natur.491..732T}
{Teanby}, N.~A., {Irwin}, P. G.~J., {Nixon}, C.~A., {et~al.} 2012, \nat, 491,
  732, \dodoi{10.1038/nature11611}

\bibitem[{{Thelen} {et~al.}(2019){Thelen}, {Nixon}, {Cordiner}, {Charnley},
  {Irwin}, \& {Kisiel}}]{2019AJ....157..219T}
{Thelen}, A.~E., {Nixon}, C.~A., {Cordiner}, M.~A., {et~al.} 2019, \aj, 157,
  219, \dodoi{10.3847/1538-3881/ab19bb}

\bibitem[{Thelen {et~al.}(2019)Thelen, Nixon, Chanover, Cordiner, Molter,
  Teanby, Irwin, Serigano, \& Charnley}]{thelen2019abundance}
Thelen, A.~E., Nixon, C., Chanover, N., {et~al.} 2019, Icarus, 319, 417,
  \dodoi{10.1016/j.icarus.2018.09.023}

\bibitem[{Vinatier {et~al.}(2020)Vinatier, Math{\'e}, B{\'e}zard, d’Ollone,
  Lebonnois, Dauphin, Flasar, Achterberg, Seignovert, Sylvestre,
  {et~al.}}]{vinatier2020temperature}
Vinatier, S., Math{\'e}, C., B{\'e}zard, B., {et~al.} 2020, Astronomy \&
  Astrophysics, 641, A116, \dodoi{10.1051/0004-6361/202038411}

\bibitem[{Virtanen {et~al.}(2020)Virtanen, Gommers, Oliphant, Haberland, Reddy,
  Cournapeau, Burovski, Peterson, Weckesser, Bright, {van der Walt}, Brett,
  Wilson, Millman, Mayorov, Nelson, Jones, Kern, Larson, Carey, Polat, Feng,
  Moore, {VanderPlas}, Laxalde, Perktold, Cimrman, Henriksen, Quintero, Harris,
  Archibald, Ribeiro, Pedregosa, {van Mulbregt}, \& {SciPy 1.0
  Contributors}}]{2020SciPy-NMeth}
Virtanen, P., Gommers, R., Oliphant, T.~E., {et~al.} 2020, Nature Methods, 17,
  261, \dodoi{10.1038/s41592-019-0686-2}

\bibitem[{{Vuitton} {et~al.}(2019){Vuitton}, {Yelle}, {Klippenstein},
  {H{\"o}rst}, \& {Lavvas}}]{2019Icar..324..120V}
{Vuitton}, V., {Yelle}, R.~V., {Klippenstein}, S.~J., {H{\"o}rst}, S.~M., \&
  {Lavvas}, P. 2019, \icarus, 324, 120, \dodoi{10.1016/j.icarus.2018.06.013}

\bibitem[{Wilson \& Atreya(2004)}]{wilson2004current}
Wilson, E.~H., \& Atreya, S. 2004, Journal of Geophysical Research: Planets,
  109, \dodoi{10.1029/2003JE002181}

\bibitem[{{Wilson} \& {Atreya}(2004)}]{2004JGRE..109.6002W}
{Wilson}, E.~H., \& {Atreya}, S.~K. 2004, Journal of Geophysical Research
  (Planets), 109, E06002, \dodoi{10.1029/2003JE002181}

\end{thebibliography}
\bibliographystyle{aasjournal}

\end{document}